\documentclass[aps,pra,twocolumn,groupedaddress,showpacs]{revtex4}
\usepackage{bm}
\usepackage{epsf}
\usepackage{amssymb}
\usepackage{graphicx}
\usepackage{rotating}
\usepackage{epsfig}
\usepackage{psfrag}
\usepackage{amsmath}
\begin{document}
\def\G{{\cal G}}
\def\F{{\cal F}}
\def\bM{{\bf M}}
\def\bN{{\bf N}}
\def\bD{{\bf D}}
\def\bSig{{\bf \Sigma}}
\def\bLam{{\bf \Lambda}}
\def\bfeta{{\bf \eta}}
\def\bc{{\bf c}}
\def\ba{{\bf a}}
\def\d{{\bf d}}
\def\bP{{\bf P}}
\def\bK{{\bf K}}
\def\bk{{\bf k}}
\def\bkn{{\bf k}_{0}}
\def\bx{{\bf x}}
\def\bz{{\bf z}}
\def\bR{{\bf R}}
\def\br{{\bf r}}
\def\bq{{\bf q}}
\def\bp{{\bf p}}
\def\bG{{\bf G}}
\def\bQ{{\bf Q}}
\def\bs{{\bf s}}
\def\E{{\bf E}}
\def\bv{{\bf v}}
\def\b0{{\bf 0}}
\def\la{\langle}
\def\ra{\rangle}
\def\Im{\mathrm {Im}\;}
\def\Re{\mathrm {Re}\;}
\def\beq{\begin{equation}}
\def\eeq{\end{equation}}
\def\bea{\begin{eqnarray}}
\def\eea{\end{eqnarray}}
\def\bdm{\begin{displaymath}}
\def\edm{\end{displaymath}}
\def\bnab{{\bf \nabla}}
\def\Tr{{\mathrm{Tr}}}
\def\bJ{{\bf J}}
\def\bU{{\bf U}}
\def\bPsi{{\bf \Psi}}
\title{Spin-polarized Fermi superfluids as Bose-Fermi mixtures}

\author{E.~Taylor}
\affiliation{Department of Physics, University of Toronto, Toronto, Ontario,
Canada, M5S 1A7,}
\author{Y.~Ohashi}
\affiliation{Faculty of Science and Technology, Keio University,
Hiyoshi, Yokohama, 223, Japan.}
\author{A.~Griffin}
\affiliation{Department of Physics, University of Toronto, Toronto, Ontario,
Canada, M5S 1A7,}
\date{\today}

\begin{abstract}
In the strong-coupling BEC region where a Feshbach resonance gives rise to tightly-bound dimer molecules, we show that a spin-polarized Fermi superfluid reduces to a simple Bose-Fermi mixture of Bose-condensed dimers and the leftover unpaired fermions.  Using a many-body functional integral formalism, the Gaussian fluctuations give rise to an induced dimer-dimer interaction mediated by the unpaired fermions, with the dimer-fermion vertex being given by the (mean-field) Born approximation.  Treating the pairing fluctuations to quartic order, we show how the action for a spin-polarized Fermi superfluid reduces to one for a Bose-Fermi mixture.  This Bose-Fermi action includes an expression for the effective dimer-unpaired fermion interaction in a spin-polarized Fermi superfluid beyond the Born approximation, in the superfluid phase at finite temperatures.  In the low-density limit, we show how this dimer-fermion interaction gives the $s$-wave scattering length $a_{BF}=1.18a_F$  ($a_F$ is the $s$-wave fermion scattering length),  a result first derived by Skorniakov and Ter-Martirosian in 1957 for three interacting fermions.  
\end{abstract}
\pacs{03.75.Kk,~03.75.Ss}
\maketitle

\section{Introduction}
Since superfluidity was first realized in ultracold trapped two-component atomic Fermi gases~\cite{Jinexpt,Grimmexpt,Ketterleexpt}, there has been increasing interest in the properties of spin-polarized Fermi superfluids~\cite{Zwierlein06,Partridge06,review}.  As in unpolarized Fermi gas superfluids, by making use of a Feshbach scattering resonance between fermions, the effective scattering length $a_{F}$ that characterizes low-energy scattering between fermions in different hyperfine states (i.e., $a_F = a_{\downarrow\uparrow}$) can be continuously tuned from negative to positive values.  We use $\uparrow,\downarrow$ to denote the two species of atomic fermions prepared in different hyperfine states.  When $a_F <0$, the system is in the BCS region and pairing can occur about the Fermi surface between fermions in different hyperfine states.  Passing through unitarity (where $|a_F| = \infty$), the scattering length becomes positive and one enters the BEC region, characterized by the disappearance of the Fermi surface (the fermion chemical potential becomes negative) and the appearance of strongly bound molecular pairs.  In the spin-polarized Fermi superfluid, there is an excess of one of the two-species of fermions: $n_{\uparrow} > n_{\downarrow}$.  In the BEC region, all of the minority species of fermions are paired up, leaving a gas of the remaining spin $\uparrow$ fermions.  Thus, in the BEC region, a spin-polarized Fermi gas should behave like a Bose-Fermi mixture of the dimer molecules and unpaired excess fermions.  In this paper, we prove that the effective action for a spin-polarized two-component Fermi superfluid in the BEC region reduces to an effective action for a Bose-Fermi mixture.

The idea that a spin-polarized Fermi superfluid reduces to a Bose-Fermi mixture in the BEC region has been explored in Refs.~\cite{Strinati06,Strinati06b,deMelo06}.  We extend those studies by considering quartic pairing fluctuations of the effective Bose action for a spin-polarized Fermi superfluid.  At this order, we show that a spin-polarized Fermi superfluid can still be described in terms of an equivalent Bose-Fermi mixture with a renormalized dimer-fermion interaction that is valid beyond the Born approximation value $U_{BF} = 2\pi(8a_F/3)/m_r$~\cite{Strinati06,Strinati06b,deMelo06}, where $m_r = 2m/3$ is the reduced mass of the molecular dimer-atomic fermion system.

Our discussion of the effective action of a spin-polarized Fermi gas is based on the functional integration treatment~\cite{Popovbook} of the BCS-BEC crossover problem developed in Refs.~\cite{deMelo, Engelbrecht}.  Extending the approach of Ref.~\cite{Ohashi05}, we expand this action in powers of pairing fluctuations about the BCS-type mean-field saddle-point up to quartic order in fluctuations.  At the level of quadratic (Gaussian) fluctuations, we derive the Bogoliubov theory of excitations of a dimer molecular condensate in the spin-polarized gas in Sec.~\ref{Bogspectrum}.  We show how the Bogoliubov excitation energy involves a dimer-dimer interaction mediated by the unpaired $\uparrow$ fermions through a Lindhard response function.  This result is compared with the Bogoliubov excitation spectrum in a Bose-Fermi mixture in Sec.~\ref{GaussianBF}.

In Sec.~\ref{quartic} we discuss the quartic fluctuations of an effective Bose action for a Bose-Fermi mixture as well as for a spin-polarized Fermi gas in the BEC limit.  Comparing the two, we prove their equivalence and derive an expression for the momentum- and energy-dependent dimer-fermion interaction in a spin-polarized.  
In order to make contact with the extensive literature on two-body dimer-fermion scattering in free space~\cite{Skorniakov57, Brodsky05, Brodsky05PRA, Gurarie06, Bedaque98}, in Sec.~\ref{TBSA}  we show how our theory reproduces the well-known result $a_{BF}=1.18a_F$ due to Skorniakov and Ter-Martirosian~\cite{Skorniakov57} for the dimer-fermion scattering length.

\section{Spin-polarized superfluid Fermi gas in the BEC limit}
\label{SPFG}

We first review the functional integral formalism for an ultracold two-component spin-polarized Fermi gas, as well as discuss some of the essential physics.  
At the low temperatures of interest, $s$-wave scattering between fermions in different hyperfine states dominates all other partial-wave amplitudes.  The $s$-wave interaction can be calculated using the pseudopotential
\bea U_{\uparrow \downarrow}(\br) = -U\delta(\br).\eea  
The partition function for the spin-polarized Fermi gas can be written as a functional integral over fermionic Grassmann fields $\psi^*,\psi$~\cite{Popovbook}:
\bea {\cal{Z}} = \int {\cal{D}}[\psi^*,\psi]e^{-S[\psi^*,\psi]},\label{Z0}\eea where 
the Euclidean (imaginary time) action is given by \bea S = \int^{\beta}_0 d\tau\left[\int d\br\;\sum_{\sigma}\psi^*_{\sigma}(x)\partial_{\tau}\psi_{\sigma}(x)  + H\right],\eea and the Hamiltonian $H$ is
\bea H&=&\int d\br\;\sum_{\sigma}\psi^*_{\sigma}(x)\left(-\frac{\bnab^2}{2m}-\mu_{\sigma}\right)\psi_{\sigma}(x)
\nonumber\\&&-U\int d\br\;\psi^*_{\uparrow}(x)\psi^*_{\downarrow}(x)\psi_{\downarrow}(x)\psi_{\uparrow}(x).\label{H}\eea Here, $x\equiv (\br,\tau)$ is a four-vector denoting the spatial coordinate $\br$ and the imaginary time variable $\tau = it$.  $\beta = 1/T$ is the inverse temperature.   We set $\hbar=k_B=1$ and the volume to be unity.  
We restrict ourselves to a uniform gas in this paper.  

The chemical potentials $\mu_{\uparrow},\mu_{\downarrow}$ in Eq.~(\ref{H}) are determined from the standard thermodynamic relation $n_{\sigma} = -(\partial\Omega/\partial\mu_{\sigma})$, where $\Omega = -T\ln {\cal{Z}}$.  A spin polarized Fermi gas (with $n_{\uparrow}\neq n_{\downarrow}$) is distinguished from an unpolarized gas by having different chemical potentials for each species: $\mu_{\uparrow}\neq \mu_{\downarrow}$.

Introducing the Bose (superfluid) pairing field $\Delta(x)$ through the Hubbard-Stratonovich transformation and integrating out the Fermi Grassmann fields, the partition function in Eq.~(\ref{Z0}) becomes:
\bea {\cal {Z}} = \int {\cal D}[\Delta, {\Delta^*}]
e^{-S_{\mathrm{eff}}[\Delta, {\Delta^*}]},\label{Bpartition}
\eea where
\bea
S_{\mathrm{eff}}[\Delta, {\Delta^*}] \equiv \int_0^\beta d\tau\int
d\br\;\frac{|\Delta(x)|^2}{U} - \mathrm{Tr}\ln [-\bG^{-1}].
\label{SeffQ} \eea The inverse of the $2\times 2$ Nambu-Gorkov BCS Green's function is
\bea \bG^{-1}(x,x') &=& \left (
\begin{array}{cc} -\partial_{\tau} + \frac{\bnab^2}{2m} + \mu_{\uparrow} &
\Delta(x)
\\
{\Delta^*}(x)
&-\partial_{\tau} - \frac{\bnab^2}{2m} - \mu_{\downarrow} 
\end{array} \right )\nonumber\\&&\times\delta(x-x'). \label{G} \eea 
This approach has been used in Ref.~\cite{deMelo06} for a spin-polarized Fermi gas.  

Expanding the action given by Eq.~(\ref{SeffQ}) up to quadratic order in fluctuations $\Lambda(x)$ about the saddle-point value $\Delta_0$ for the Bose-pairing field, and Fourier-transforming the resulting expression, the 
BCS mean-field $S^{(0)}$ and Gaussian fluctuation $S^{(2)}$ contributions are given by~\cite{deMelo06}
\bea S^{(0)} &=& \beta\frac{\Delta^{2}_0}{U} - \sum_{k}\mathrm{tr}\ln
[-\bG_0^{-1}(k)]
\label{S0k}\eea
and 
\bea S^{(2)} &=& \sum_q\frac{|\Lambda_q|^2}{U}\nonumber\\&&+
\frac{1}{2\beta }\sum_{k,q}\mathrm{tr}[\bG_0(k)\bSig(-q)
\bG_0(k+q)\bSig(q)]
\nonumber\\ &\equiv&
\frac{1}{2}\sum_q\bLam^{\dagger} \bM \bLam.
\label{S2kb} \eea Here,
\bea \bG_0(k) &\equiv& \frac{1}{(ik_n - E^+_{\bk})(ik_n + E^-_{\bk})}\nonumber\\ &&\times
\left (
\begin{array}{cc} ik_n + \xi_{\bk,\downarrow} &
-\Delta_0
\\
-\Delta_0
&ik_n - \xi_{\bk,\uparrow} 
\end{array} \right )
\label{G0Q} \eea
is the BCS $2\times 2$ matrix Green's
function for the spin-polarized Fermi superfluid, and
the fluctuation self-energy is defined as
\bea \bSig(q) = 
\left (
\begin{array}{cc} 0 &
\Lambda_q
\\
\Lambda^*_{-q}
&0
\end{array} \right ),\label{bSig}\eea
and $\xi_{\bk,\sigma} \equiv \bk^2/2m - \mu_{\sigma}$.  The poles of $\bG_0(k)$ are given by \bea E^{\pm}_{\bk} = 
 \sqrt{(\xi^+_{\bk})^2 + \Delta^2_0}\pm \xi^{-}_{\bk},\label{Epm} \eea  where $\xi^{\pm}_{\bk} = (\xi_{\bk,\uparrow} \pm \xi_{\bk,\downarrow})/2$.  The Green's function in Eq.~(\ref{G0Q}) describes the BCS excitation spectrum~\cite{review,deMelo06} in a spin-polarized Fermi superfluid.  It reduces to the standard result when $\mu_{\uparrow}=\mu_{\downarrow}$, in which case $\xi^+_{\bk} = \bk^2/2m - \mu$ and $\xi^-_{\bk} = 0$.  In the last line of
Eq.~(\ref{S2kb}), we have defined the spinor 
$\bLam^{\dagger} \equiv (\Lambda^*_q, \Lambda_{-q})$.
Also, $k \equiv (\bk,ik_n)$ and $q \equiv (\bq,iq_m)$ are four-vectors denoting the wave vectors $\bk$ and $\bq$ as well as the 
Fermi and Bose Matsubara frequencies, $k_n=(2n+1)\pi/\beta$ and $q_m= 2m\pi/\beta$, respectively.  Throughout this paper, $k_n$ shall refer to a Fermi frequency and $q_m$ a Bose frequency.

The matrix elements of the inverse $2\times2$ matrix pair fluctuation
propagator $\bM$ defined in Eq.~(\ref{S2kb}) are given by~\cite{Engelbrecht,Taylor06}
\bea \lefteqn{M_{11}(q) =
M_{22}(-q)=}&&\nonumber\\&&\frac{1}{U} +
\frac{1}{\beta}\sum_k G_{0,11}(k+q)G_{0,22}(k)
\label{m11Q}\eea
and
\bea M_{12}(q)=M_{21}(q)=\frac{1}{\beta}\sum_k
G_{0,12}(k+q)G_{0,12}(k).\label{m12Q}\eea 
Here, $G_{0,ij}$ denotes the $ij$-th element of the
Green's function defined by Eq.~(\ref{G0Q}). 

The number of fermions in state $\sigma$ is given by $n_{\sigma} = -(\partial\Omega/\partial\mu_{\sigma}$), where the thermodynamic potential is~\cite{Taylor06} \bea \Omega \simeq \frac{\Delta^{2}_0}{U} - \frac{1}{\beta}\sum_{k}\mathrm{tr}\ln
[-\bG_0^{-1}(k)]+\frac{1}{2\beta}\sum_q\ln\mathrm{det}\bM(q).\nonumber\\ \eea 
In the BEC limit [$(k_Fa_F)^{-1}\gg 1$], the chemical potential for the minority species is large and negative, roughly corresponding to the binding energy of a dimer molecule~\cite{deMelo06,Strinati06,Strinati06b,Liu06}.  Using this fact, one can show in the BEC limit and above the superfluid transition temperature $T_c$ that (see also the related discussion in Ref.~\cite{Liu06})
\bea n_{\sigma} = \sum_{\bk}f(\xi_{\bk,\sigma}) + \sum_{\bq}n_B(\xi_{B,\bq}),\label{nsig}\eea
where $\xi_{B,\bq} \equiv \bq^2/2M - \mu_{B}$ with the boson chemical potential $\mu_B \equiv \mu_{\downarrow} +  \mu_{\uparrow}+ |E_b|$~\cite{Strinati06,deMelo06} and molecular mass $M=2m$.  Here, $f(x) = (e^{\beta x}+1)^{-1}$ and $n_B(x) = (e^{\beta x}-1)^{-1}$ are the Fermi and Bose thermal distribution functions, respectively.  The Fermi distribution function describes the unpaired free fermions while the Bose distribution function describes the fermions in bound states.  
Deep in the BEC region, there will be no unpaired $\downarrow$ fermions, and thus $f(\xi_{\bk,\downarrow})\rightarrow 0$.  This means that the number of excess unpaired $\uparrow$ fermions is given by [using Eq.~(\ref{nsig})]
\bea \delta n_F \equiv n_{\uparrow}-n_{\downarrow} &=& \sum_{\bk}\left[f(\xi_{\bk,\uparrow}) - f(\xi_{\bk,\downarrow})\right]\nonumber\\&\simeq& \sum_{\bk}f(\xi_{\bk,\uparrow}). \label{deltan2}\eea
This shows that the chemical potential $\mu_{\uparrow}$ for the majority species determines the number of unpaired excess fermions.  One can think of $\mu_{\uparrow}$ as being roughly equal to the Fermi energy of a gas of unpaired fermions of density $\delta n_F$, i.e., $\mu_{\uparrow} \approx (6\pi^2 \delta n_F)^{2/3}/2m$.  Deep in the BEC region, the dimer-unpaired fermion interactions are weak and  Eq.~(\ref{deltan2}) will also be valid in the superfluid phase.     


In the strong-coupling BEC limit, $\Delta_0$ is much smaller than the dimer binding energy~\cite{Engelbrecht} and consequently, $|\mu_{\downarrow}|\sim 1/ma^2_F = |E_b| \gg \Delta_0$.  In this limit, the two quasiparticle branches in Eq.~(\ref{Epm}) can be approximated by 
\bea E^{+}_{\bk} \simeq \frac{\bk^2}{2m} - \mu_{\uparrow} + g(\bk)\label{Ep}\eea
and
\bea E^{-}_{\bk} &\simeq& \frac{\bk^2}{2m} -\mu_{\downarrow}+ g(\bk)\nonumber\\
&\simeq& \frac{\bk^2}{2m} + |E_b| + g(\bk),\label{Em} \eea
where
\bea  g(\bk)\equiv \frac{\Delta^2_0}{\bk^2/m + |\mu_{\downarrow}|}.\label{term}\eea
In Sec.~\ref{quartic}, we will show that $g(\bk)$ is the low-density limit of the dimer-fermion interaction for a fermion of momentum $\bk$ interacting with a dimer boson, multiplied by the condensate density of dimers.  Thus, $g(\bk)$ is the expected self-energy correction for the fermions due to interactions with the Bose condensate of dimer molecules.  Hence, $E^+_{\bk}$ in Eq.~(\ref{Ep}) is precisely the spectrum one would expect for an unpaired fermion interacting with a Bose-condensate of dimer molecules.  The second fermionic branch, given by Eq.~(\ref{Em}), involves the break-up of a dimer pair and this excitation is frozen out in the BEC limit, where the dimer binding energy $|E_b|\gg\Delta_0,T_c$ is very large.


\section{Bogoliubov spectrum in the Gaussian approximation}
\label{Bogspectrum}

In this Section, we will derive the Bogoliubov excitation energy in a spin-polarized Fermi gas in the BEC region using the results of Sec.~\ref{SPFG}.  
Summing over the fermion Matsubara frequencies in Eqs.~(\ref{m11Q}) and
(\ref{m12Q}), the matrix elements of the inverse matrix propagator
for pair fluctuations are given by (see Refs.~\cite{Engelbrecht,Taylor06})
\bea \lefteqn{M_{11}(q) = M_{22}(-q) =}&&\nonumber\\&& \frac{1}{U}
+\sum_{\bk}\Bigg[\left(f^+_{\bk}
- f^-_{\bk + \bq}\right)
\frac{v^2_{\bk}v^2_{\bk
+ \bq}}{iq_m + E_{\bk} + E_{\bk +
\bq}}\nonumber\\
&&+\left(f^-_{\bk} - f^+_{\bk + \bq}\right) \frac{u^2_{\bk}u^2_{\bk +
\bq}}{iq_m - E_{\bk} - E_{\bk +
\bq}}\nonumber\\
&&+\left(f^+_{\bk} - f^+_{\bk + \bq}\right) \frac{v^2_{\bk}u^2_{\bk +
\bq}}{iq_m + E_{\bk} - E_{\bk +
\bq}}\nonumber\\
&&+\left(f^-_{\bk} - f^-_{\bk + \bq}\right) \frac{u^2_{\bk}v^2_{\bk +
\bq}}{iq_m - E_{\bk} + E_{\bk + \bq}}
\Bigg] \label{m11b}\eea
and
\bea \lefteqn{M_{12}(q) =  M_{21}(q)=}&&\nonumber\\ &&\sum_{\bk}\Bigg[\left(f^-_{\bk+
\bq} -
f^+_{\bk}\right) \frac{u_{\bk}v_{\bk}u_{\bk + \bq}v_{\bk + \bq}}{iq_m +E_{\bk} + E_{\bk + \bq}}\nonumber\\
&&+\left(f^+_{\bk+\bq} - f^-_{\bk}\right) \frac{u_{\bk}v_{\bk}u_{\bk +
\bq}v_{\bk + \bq}}{iq_m - E_{\bk} -E_{\bk +
\bq}}\nonumber\\
&&+\left(f^+_{\bk} - f^+_{\bk + \bq}\right) \frac{u_{\bk}v_{\bk}u_{\bk +
\bq}v_{\bk + \bq}}{iq_m + E_{\bk} - E_{\bk +
\bq}}\nonumber\\
&&+\left(f^-_{\bk} - f^-_{\bk + \bq}\right) \frac{u_{\bk}v_{\bk}u_{\bk +
\bq}v_{\bk + \bq}}{iq_m - E_{\bk} + E_{\bk +
\bq}}\Bigg], \label{m12b}\eea
where
\bea f^{\pm}_{\bp} \equiv f\left(\pm E^{\pm}_{\bp}\right) \eea
are
the Fermi
distribution functions and  
\bea E_{\bp} \equiv \sqrt{(\xi^+_{\bp})^2 + \Delta^2_0} = E^+_{\bp} -\xi^-_{\bp}.\eea
Here, $u_{\bp} = \sqrt{(1 + \xi^+_{\bp}/E_{\bp})/2}$ and  $v_{\bp}
= \sqrt{(1 -
\xi^+_{\bp}/E_{\bp})/2}$ are the Bogoliubov quasiparticle
amplitudes for a spin-polarized Fermi superfluid.  We note that $E_{\bp}$ is not the physical quasiparticle dispersion (given by $E^{\pm}_{\bp}$), except when the polarization vanishes ($\xi^-_{\bp}= 0;\;\xi^+_{\bp}=\xi_{\bp}$).  

The processes described by the first two lines (on the right-hand side) in Eqs.~(\ref{m11b}) and (\ref{m12b}) correspond to pair creation/destruction of two excitations $\pm (E_{\bk}+E_{\bk+\bq})$, while the last two terms describe creation/destruction of particle-hole excitations $\pm (E_{\bk}-E_{\bk+\bq})$.  Following the usual approach for unpolarized Fermi superfluids~\cite{Engelbrecht}, in the BEC limit, we expand the first and second lines of Eqs.~(\ref{m11b}) and (\ref{m12b}) in powers of $q=(\bq,iq_m)$.  For an unpolarized gas ($\mu_{\downarrow}=\mu_{\uparrow} \simeq -|E_b|/2$), $f^+\rightarrow 0$ and $f^-\rightarrow 1$ since $T \ll |E_b|$ in the BEC region (recall that $T_c\sim 0.2\epsilon_F$ in the BEC region while $|E_b| = 2(k_Fa_F)^{-2}\epsilon_F$ is much larger when $(k_Fa_F)^{-1}\gg 1$).  Thus, the third and fourth lines in  Eqs.~(\ref{m11b}) and (\ref{m12b}), describing particle-hole excitations, vanish.  However, for the polarized superfluid, using Eq.~(\ref{Em}) we see that in the BEC limit we still have that $f^- \rightarrow 1$ since the chemical potential for the minority species is large and negative.  On the other hand, $\mu_{\uparrow}\approx \epsilon_F(\delta n_F)$ is \textit{positive}, where $\epsilon_F(\delta n_F)$ is the Fermi energy of the excess fermions, with density $\delta n_F = n_{\uparrow}-n_{\downarrow}$.  Thus, \bea f^+_{\bp}&\equiv& f(E^+_{\bp})\nonumber\\&\simeq& f(\xi_{\bp,\uparrow} + g(\bp))\nonumber\\
&\equiv& f(\xi^{\prime}_{\bp,\uparrow}),\label{xip}\eea does \textit{not} vanish in the BEC limit.  This means that for a polarized Fermi superfluid in the BEC limit, the particle-hole terms given by the third lines in both Eqs.~(\ref{m11b}) and (\ref{m12b}) do not vanish.  We will show that these terms (which we denote as $R_{11}(q)$ and $R_{12}(q)$, respectively) can be written in terms of a fermionic particle-hole density response function in the BEC limit [see Eq.~(\ref{R11R12})].  

Following the previous discussion, in the BEC limit, the matrix elements of the inverse pair fluctuation propagator in Eqs.~(\ref{m11b}) and (\ref{m12b}) are approximated by \bea M_{11} \simeq A + B\bq^2 - C(iq_m) + R_{11}(q) \label{m11f}\eea and \bea M_{12} \simeq A + D\bq^2 + R_{12}(q),\label{m12f}\eea  where the expansion coefficients $(A,B,C$, and $D$) describe the small-$q$ expansion of the first two lines in Eqs.~(\ref{m11b}) and (\ref{m12b}). To leading order in $(\Delta_0/|\mu_{\downarrow}|)^2$, these coefficients reduce to (see Refs.~\cite{Engelbrecht} and \cite{Taylor06} for similar calculations)
\bea A = \sum_{\bk}\frac{\Delta^2_0}{4(\xi^+_{\bk})^3}\left(1-f(\xi'_{\bk,\uparrow})\right),\label{A0}\eea
\bea B = \sum_{\bk}\left(1-f(\xi'_{\bk,\uparrow})\right)\left[\frac{1}{8m(\xi^+_{\bk})^2} - \frac{(\bk\cdot\hat{\bq}/m)^2}{8(\xi^+_{\bk})^3}\right],\eea
and
\bea C = \sum_{\bk}\frac{1}{4(\xi^+_{\bk})^2}\left(1-f(\xi'_{\bk,\uparrow})\right),\label{C0}\eea
where $\hat{\bq} =\bq/|\bq|$.  $D\sim\Delta^2_0a^5_F$ is vanishingly small in the BEC limit~\cite{Taylor06}.  

The mean-field BCS gap equation follows from the requirement that the saddle-point action $S^{(0)}$ be stationary with respect to fluctuations $\Lambda(x)$ about $\Delta_0$: $\partial S^{(0)}/\partial\Delta_0 = 0$.  This gives the usual expression
\bea \frac{\Delta_0}{U} = \frac{1}{\beta}\sum_kG_{0,12}(k).\label{gap0}\eea 
In the BEC limit, the off-diagonal BCS Green's functions in Eq.~(\ref{G0Q}) are given by
\bea G_{0,12}(k) = G_{0,21}(k)\simeq -\Delta_0\tilde{G}_{0,\uparrow}(k)\tilde{G}_{0,\downarrow}(k),\label{G12}\eea
where we have introduced the normal single-particle Green's functions
\bea G_{0,\downarrow}(k) = \frac{1}{ik_n + \xi_{\bk,\downarrow}},\:\:\: G_{0,\uparrow}(k) = \frac{1}{ik_n - \xi_{\bk,\uparrow}}.
\label{Gup}\eea
Using Eq.~(\ref{G12}), Eq.~(\ref{gap0}) reduces to
\bea \frac{1}{U}&=&-\frac{1}{\beta}\sum_k \tilde{G}_{0,\uparrow}(k)\tilde{G}_{0,\downarrow}(k)\nonumber\\&=&
\sum_{\bk}\frac{1-f(\xi'_{\bk,\uparrow})}{\xi'_{\bk,\uparrow} + \xi'_{\bk,\downarrow}}.\label{gap}\eea 
Introducing the standard regularized two-body potential~\cite{deMelo} \bea \frac{1}{U} = -\frac{m}{4\pi a_F} + \sum_{\bk}\frac{m}{\bk^2},\label{LS}\eea we can replace $1/U$ in Eq.~(\ref{gap}) in terms of $a_F$.   In order to solve Eq.~(\ref{gap}) analytically, we use the approximation $g(\bk)\simeq g(\b0) = \Delta^2_0/|\mu_{\downarrow}|$, valid at the mean-field level for low densities of the unpaired $\uparrow$ fermions since $\bk^2_F/2m \simeq \mu_{\uparrow} \ll |\mu_{\downarrow}|$, where $k_F \equiv (6\pi^2\delta n_F)^{1/3}$ is the Fermi wave vector for the unpaired $\uparrow$ fermions.  We find
\bea \lefteqn{\sqrt{|\mu_{\downarrow}| - \mu_{\uparrow} + 2g(\b0)} =}&&\nonumber\\&&\frac{1}{\sqrt{m}a_F}\Bigg[1 - \frac{4\pi a_F}{m}\sum_{\bk}\frac{f(\xi'_{\bk,\uparrow})}{\bk^2/m + |\mu_{\downarrow}| - \mu_{\uparrow}}\Bigg].\label{gap1}\eea
Since $f(\xi'_{\bk,\uparrow})\rightarrow 0$ for $k>k_F\approx \sqrt{2m\mu_{\uparrow}}$ and $|\mu_{\downarrow}| \gg \mu_{\uparrow}$, we have
\bea \sum_{\bk}\frac{f(\xi'_{\bk,\uparrow})}{\bk^2/m + |\mu_{\downarrow}|-\mu_{\uparrow}} \approx \frac{1}{|\mu_{\downarrow}|}\sum_{\bk}f(\xi'_{\bk,\uparrow}) \equiv \frac{1}{|\mu_{\downarrow}|}
\delta n_F,\nonumber\\ \label{approx}\eea and thus Eq.~(\ref{gap1}) reduces to
\bea |\mu_{\downarrow}| - \mu_{\uparrow} \approx |E_b|\left(1 - 8\pi(\delta n_Fa^3_F) - 2\frac{\Delta^2_0}{|E_b|^2}\right).\label{gap2}\eea  Recall that $\delta n_F$ is the density of the unpaired excess $\uparrow$ fermions.  

The corrections on the right-hand side of Eq.~(\ref{gap2}), due to a finite density of unpaired fermions as well as a finite condensate density, are both higher-order corrections (${\cal{O}}[\delta n_F a^3_F]$ and ${\cal{O}}[n_F a^3_F]$, respectively~\cite{note}).  In the BEC limit, we ignore these corrections.  We note that $|E_b|(\delta n_Fa^3_F) = (k^2_F/2m)(2k_Fa_F) \approx \mu_{\uparrow}(2k_Fa_F)$.  Thus, the contributions we drop from Eq.~(\ref{gap2}) are much smaller than $\mu_{\uparrow}$ in the BEC limit and the gap equation reduces to \bea |\mu_{\downarrow}| - \mu_{\uparrow}= \frac{1}{ma^2_F}\left(1+{\cal{O}}[\delta n_Fa^3_F]\right).\label{gap3}\eea In Eqs.~(\ref{A0})-(\ref{C0}), the terms containing the Fermi thermal factor $f(\xi'_{\bk,\uparrow})$ are also ${\cal{O}}[\delta n_Fa^3_F]$, as can be shown by using the approximation in Eq.~(\ref{approx}).  Using Eq.~(\ref{gap3}), the expansion coefficients reduce to
\bea A = \frac{\Delta^2_0a^3_F m^3}{16\pi}\left(1+{\cal{O}}[\delta n_Fa^3_F]\right),\label{A}\eea
\bea B = \frac{ma_F}{32\pi}\left(1+{\cal{O}}[\delta n_Fa^3_F]\right),\label{B}\eea
and
\bea C = \frac{m^2a_F}{8\pi}\left(1+{\cal{O}}[\delta n_Fa^3_F]\right).\label{C}\eea

We now turn to the evaluation of the $R_{11}$ and $R_{12}$ terms in Eqs.~(\ref{m11f}) and (\ref{m12f}).  Using the expansion \bea E_{\bk}&\simeq& \xi^+_{\bk} + \Delta^2_0/2\xi^+_{\bk}\nonumber\\&\simeq& \xi^+_{\bk} +g(\bk)\eea in the third lines of Eqs.~(\ref{m11b}) and (\ref{m12b}), we obtain 
\bea R_{11}(q)\simeq R_{12}(q) \simeq \sum_{\bk}\frac{g^2(\bk)}{\Delta^2_0}\frac{f(\xi'_{\bk,\uparrow}) - f(\xi'_{\bk+\bq,\uparrow})}{iq_m +\xi'_{\bk,\uparrow} -  \xi'_{\bk+\bq,\uparrow}}.\label{R11R12}\eea  As we shall show later, this term is of the order \bea \Delta^2_0a^3_F m^3(\delta n_Fa^3_F)^{1/3}.\label{magnitude}\eea  Thus, comparing with the finite-density corrections to the coefficients given in Eqs.~(\ref{A})-(\ref{C}), we see that this is the leading-order finite-density correction.

In summary, for small energy and momentum $q$, the matrix elements of the inverse pair fluctuation propagator $\bM$ defined in Eqs.~(\ref{m11b}) and (\ref{m12b}) are given by
\bea M_{11} &=& A + B\bq^2 - C(iq_m) \nonumber\\&&+ \sum_{\bk}\frac{g^2(\bk)}{\Delta^2_0}\frac{f(\xi'_{\bk,\uparrow}) - f(\xi'_{\bk+\bq,\uparrow})}{iq_m+\xi'_{\bk,\uparrow} -  \xi'_{\bk+\bq,\uparrow}}
\nonumber\\&=&\frac{m^2a_F}{8\pi}\Bigg[-iq_m + \frac{\bq^2}{4m} + \frac{\Delta^2_0a^2_Fm}{2} \nonumber\\&&+
\frac{1}{n_c}\sum_{\bk}g^2(\bk)\frac{f(\xi'_{\bk,\uparrow}) - f(\xi'_{\bk+\bq,\uparrow})}{iq_m+\xi'_{\bk,\uparrow} -  \xi'_{\bk+\bq,\uparrow}}\Bigg]\nonumber\\
&=&\frac{m^2a_F}{8\pi}\left[-iq_m + \frac{\bq^2}{2M} + n_c\tilde{U}^{(0)}_{BB}(q)\right]\label{m11c}\eea
and 
\bea M_{12}&=& A + \sum_{\bk}\frac{g^2(\bk)}{\Delta^2_0}\frac{f(\xi'_{\bk,\uparrow}) - f(\xi'_{\bk+\bq,\uparrow})}{iq_m+\xi'_{\bk,\uparrow} -  \xi'_{\bk+\bq,\uparrow}}\nonumber\\
&=&\frac{m^2a_F}{8\pi}\left[n_c\tilde{U}^{(0)}_{BB}(q)\right], \label{m12c}\eea
where we have defined the effective interaction
\bea \tilde{U}^{(0)}_{BB}(\bq,iq_m) \equiv U^{(0)}_{BB} + \sum_{\bk}\frac{g^2(\bk)}{n^2_c}\frac{f(\xi'_{\bk,\uparrow}) - f(\xi'_{\bk+\bq,\uparrow})}{iq_m+\xi'_{\bk,\uparrow} -  \xi'_{\bk+\bq,\uparrow}}.\nonumber\\ \label{UBBtilde}\eea
The direct dimer-dimer interaction is  \bea U^{(0)}_{BB} = \frac{4\pi(2a_F)}{M}, \label{UBB0}\eea where the molecular scattering length is given by the Born approximation $a_{BB} = 2a_F$~\cite{deMelo}. 
The condensate density $n_c$ of dimer molecules which appears in Eq.~(\ref{UBBtilde}) is given in the BEC limit by~\cite{Fukushima} \bea n_c(T) = \frac{\Delta^2_0(T) m^2 a_F}{8\pi}. \label{nc}\eea

Substituting the results in Eqs.~(\ref{m11c}) and (\ref{m12c}) into Eq.~(\ref{S2kb}) and defining the Bose fluctuation field
\bea c_q \equiv \sqrt{\frac{m^2a_F}{8\pi}}\Lambda_q,\label{cq}\eea the action describing Gaussian fluctuations past the BCS mean-field is given by 
\bea  S^{(2)}&=& -\frac{1}{2}{\sum_q}'\bc^{\dagger}_q \bD^{-1}(q)\bc_{q},\label{S2sp}\eea where
$\bc^{\dagger}_{q}\equiv (c^*_q,c_{-q})$ and $\bD^{-1}$ is the $2\times 2$ inverse matrix propagator for the Bogoliubov excitations of a dimer condensate in a spin-polarized superfluid:\bea \lefteqn{-\bD^{-1}(q)\equiv}&&\nonumber\\&&
\!\!\!\!\left (
\begin{array}{cc} -iq_m + \frac{\bq^2}{2M} +n_c\tilde{U}^{(0)}_{BB}(q) &
n_c\tilde{U}^{(0)}_{BB}(q)
\\
n_c\tilde{U}^{(0)}_{BB}(q)
&iq_m + \frac{\bq^2}{2M} + n_c\tilde{U}^{(0)}_{BB}(q)
\end{array} \right).\nonumber\\ \label{Dtilde}\eea
Equation~(\ref{Dtilde}) is identical to the propagator for Bogoliubov excitations in an unpolarized Fermi superfluid in the BEC limit~\cite{Engelbrecht, Taylor06}, apart from the renormalized interaction $\tilde{U}^{(0)}_{BB}(q)$ defined in Eq.~(\ref{UBBtilde}).

We now discuss the Bogoliubov excitations of the dimer condensate which are described by the propagator defined in Eq.~(\ref{Dtilde}).  If we replace $g(\bk)$ by its long-wavelength value $g(\b0) = \Delta^2_0/|\mu_{\downarrow}|$, $\tilde{U}^{(0)}_{BB}$ in Eq.~(\ref{UBBtilde}) becomes
\bea \tilde{U}^{(0)}_{BB}(q) \approx U^{(0)}_{BB} + \left(\frac{8\pi a_F}{m}\right)^2\chi(q,T),\label{Utilde}\eea
where 
\bea \chi(\bq,iq_m;T) \equiv \sum_{\bk}\frac{f(\xi'_{\bk,\uparrow}) - f(\xi'_{\bk+\bq,\uparrow})}{iq_m+\xi'_{\bk,\uparrow} -  \xi'_{\bk+\bq,\uparrow}},\label{chifull}\eea
is the Lindhard response function~\cite{Mahan,pethickbook,Buchler} for the excess $\uparrow$ fermions.  This function (see, for example, Ref.~\cite{Mahan}) describes the induced fermionic density fluctuation in response to a density fluctuation in the bosons with energy $iq_m$ and momentum $\bq$.  The fermion density fluctuation acts back on the bosons to give rise to a fermion-mediated boson-boson interaction.  

At low temperatures ($T\ll T_F$), one finds Eq.~(\ref{chifull}) gives \bea \chi(0,0)\simeq -N(\epsilon_F) \simeq -\frac{mk_F}{2\pi^2},\label{chi}\eea where $N(\epsilon)$ is the fermionic density of states and  $\epsilon_F = k^2_F/2m$ is the Fermi energy of the unpaired $\uparrow$ fermions.  In Eq.~(\ref{chi}) we have used the density of states $N(\epsilon_F) = 3\delta n_F/2\epsilon_F$ of an ideal single-component Fermi gas of density $\delta n_F$.  Using $k_F = (6\pi^2\delta n_F)^{1/3}$ in Eq.~(\ref{chi}) leads to the result given in Eq.~(\ref{magnitude}) [giving the magnitude of $g^2(0)\chi(0,0)/\Delta^2_0$].  We thus see that this correction in Eq.~(\ref{Utilde}) due to the effects of the unpaired excess fermions is much larger than the corrections we dropped in Eqs.~(\ref{A})-(\ref{C}).  These higher-order terms due to a finite density of unpaired fermions would renormalize the parameters ($n_c, M, U^{(0)}_{BB}$) appearing in the Bogoliubov propagator.  However, we do not discuss this extension here.

The long-wavelength bosonic spectrum of a spin-polarized Fermi superfluid in the BEC limit [found from the solution of $\mathrm{det}\bD^{-1}(\bq,\omega_{\bq})=0$ given by Eq.~(\ref{Dtilde})], are the usual Bogoliubov excitations,
\bea \omega_{\bq} = \sqrt{(c\bq)^2 + \Big(\frac{\bq^2}{2M}\Big)^2}, \eea
where the renormalized sound velocity $c$ is defined by
\bea c^2 = \frac{\tilde{U}^{(0)}_{BB}(\b0)n_c}{M}.\label{sound}\eea  Since the induced interaction $(8\pi a_F/m)^2\chi(0,0) \simeq - (8\pi a_F/m)^2mk_F/2\pi^2$ is attractive, the sound velocity becomes smaller as the density $\delta n_F = k^3_F/6\pi^2$ of the excess unpaired fermions is increased.  Above a critical value of the density corresponding to $c=0$, the sound velocity becomes imaginary, signalling an instability towards phase-segregation of the bound dimers and unpaired fermions.  
Using Eqs.~(\ref{Utilde}), (\ref{chi}), and (\ref{sound}), one can show that the sound velocity is real as long as
\bea \delta n_F^{1/3} \le \frac{(6\pi^2)^{2/3}}{3m}\frac{U^{(0)}_{BB}}{(8\pi a_F/m)^2}.\label{conditionPS}\eea
Thus, Eq.~(\ref{conditionPS}) gives the stability condition for the uniform superfluid phase of a spin-polarized Fermi gas in the BEC region [see also Eq.~(\ref{conditionPS2}) for further discussion].  

Summarizing the results of this Section, we have given an explicit expression for the spectrum of Bogoliubov excitations of the dimer condensate in the BEC limit of a spin-polarized Fermi superfluid.  Since the renormalized phonon sound speed becomes imaginary above a critical density of the unpaired $\uparrow$ fermions, we showed how
the Bogoliubov spectrum also gives the condition for the system to become unstable towards phase segregation.  In order to better understand the results given here, in Sec.~\ref{GaussianBF} we discuss the Gaussian fluctuations of a Bose-Fermi mixture and show that they are equivalent to the results of this Section for a spin-polarized Fermi superfluid.

\section{Gaussian approximation for an effective Bose-Fermi theory}
\label{GaussianBF}
In order to understand our expression for the effective dimer-dimer interaction in Eq.~(\ref{UBBtilde}) that arises in the Bogoliubov excitation spectrum for a spin-polarized Fermi superfluid in the BEC limit, in this Section we consider the Gaussian fluctuations of a Bose-Fermi mixture.  We show that at the Gaussian level, a spin-polarized Fermi superfluid in the BEC limit can be modelled as a Bose-Fermi mixture and identify the effective dimer-unpaired fermion interaction (within the Born approximation).  

We introduce the following action for a Bose-Fermi mixture:
\bea S[a,c] &=&
\sum_{k}\left(-ik_n
+\frac{\bk^2}{2m}-\mu_{F}\right)a^*_{k,\uparrow}a_{k,\uparrow}\nonumber\\&&
\!\!\!\!+\sum_{q}\left(-iq_m +\frac{\bq^2}{2M} -
\mu_B\right)c^*_{q}c_{q}\nonumber\\
&&\!\!\!\!+\frac{1}{\beta}\sum_{k,k',P}\!\!
U_{BF}(k,k',P)
c^*_{P-k}c_{P-k'}a^*_{k,\uparrow}a_{k',\uparrow}
\nonumber\\&&\!\!\!\!+\frac{1}{2\beta}\!
\sum_{q,q',K}\!\!U_{BB}(q,q',K)c^*_{\frac{K}{2}+q'}c^*_{\frac{K}{2}-q'}c_{\frac{K}{2}-q}c_{\frac{K}{2}+q}
.\nonumber\\ \label{Seff}\eea
Here, $P\equiv (\bP,iP_n)$ and $P_n$ is a Fermi Matsubara frequency representing the centre-of-mass degree of freedom of a boson-fermion pair.  As with before, $k_n,k'_n$ are also Fermi frequencies, while $K_m$ and $q_m$ are Bose frequencies.  
Equation (\ref{Seff}) describes a system of bosons ($c$) of mass $M$ and fermions ($a$) of mass $m$ interacting via the potential $U_{BF}$, and bosons interacting with each other through $U_{BB}$.  $c$ is a complex number Bose field, while $a$ is a Fermi Grassmann field.  

$\mu_B$ and $\mu_F$ refer to the chemical potentials of the boson and fermions, respectively.  In contrast to the spin-polarized gas in Sec.~\ref{SPFG}, only one species of fermions with chemical potential $\mu_F$ appears in Eq.~(\ref{Seff}).  This chemical potential is the same as $\mu_{\uparrow}$ in the spin-polarized gas, however, being determined by Eq.~(\ref{deltan2}). 


Integrating out the Fermi Grassmann fields from the partition function ${\cal{Z}} = \int{\cal{D}}[a,c]e^{-S[a,c]}$, we obtain an effective Bose action which can be compared directly with the effective action for the bosonic fluctuations of a spin-polarized Fermi superfluid.     
The model Bose-Fermi action given in Eq.~(\ref{Seff}) is bilinear in the Fermi Grassmann fields $a,a^*$, and the Gaussian integral can be evaluated in the usual way to give
\bea S_{\mathrm{eff}}[c] = S_B - \mathrm{Tr}\ln[- A],\label{SeffB}\eea
where  
\bea
S_B &\equiv& \sum_{q}\left(-iq_m +\xi_{B,\bq}\right)c^*_{q}c_{q}\nonumber\\&&\!\!\!\!+
\frac{1}{2\beta}\sum_{q,q',K}U_{BB}(q,q',K)c^*_{\frac{K}{2}+q'}c^*_{\frac{K}{2}-q'}c_{\frac{K}{2}-q}c_{\frac{K}{2}+q} \nonumber\\ \label{SB} \eea
and \bea A(k,k') &\equiv& (ik_n - \xi_{F,\bk})\delta_{k,k'} \nonumber\\&&- \frac{1}{\beta}\sum_PU_{BF}(k,k',P)
c^*_{P-k}c_{P-k'},\eea
with $\xi_{B,\bq} \equiv \bq^2/2M - \mu_B$ and $\xi_{F,\bk} \equiv \bk^2/2m - \mu_F$.

Treating fluctuations of the Bose condensate as small, we 
apply the Bogoliubov shift $c_q\rightarrow \sqrt{n_c\beta} + c_q$ to $A(k,k')$ and expand about its saddle point value $A_0(k)$: \bea A(k,k') = A_0(k) + \Lambda^{(1)}(k,k') +  \Lambda^{(2)}(k,k'),\eea where
\bea A_0(k) &=& ik_n - \xi_{F,\bk} - n_cU_{BF}(k,k,k)
\nonumber\\&\equiv& ik_n - \xi'_{F,\bk},
\label{A_02}\eea
\bea \lefteqn{\Lambda^{(1)}(k,k') =-\sqrt{\frac{n_c}{\beta}}}&&\nonumber\\&&\times\left[U_{BF}(k,k',k)c_{k-k'}+ U_{BF}(k,k',k')c^*_{k'-k}\right], \eea and
\bea \Lambda^{(2)}(k,k') = -\frac{1}{\beta}
\sum_P U_{BF}(k,k',P)c^*_{P-k}c_{P-k'}.\label{Lambda2}\eea
Using this expansion in Eq.~(\ref{SeffB}), the saddle-point action is given by
\bea S^{(0)} = -\mu_B n_c \beta + \frac{U_{BB}(0,0,0)n^2_c\beta}{2} -
\mathrm{Tr}\ln[-A_0]. \label{S0}\eea
The Gaussian fluctuations are described by
\bea S^{(2)} = S^{(2)}_B - \mathrm{Tr}[A^{-1}_0\Lambda^{(2)}] + \frac{1}{2} \mathrm{Tr}[(A^{-1}_0\Lambda^{(1)})^2], \label{S2}\eea
where $S^{(2)}_B$ gives the Gaussian fluctuation terms from $S_B$ in Eq.~(\ref{SB}). 

We now show that Eq.~(\ref{S2}) is equivalent to the analogous terms in Eq.~(\ref{S2sp}) for a spin-polarized Fermi superfluid.  
Performing the frequency sums in the two terms in Eq.~(\ref{S2}) and replacing terms like $U_{BF}(k,k+q,k;ik_n=\xi_{F,\bk})$ by its free-space ($\mu_{F}=0$), long-wavelength, static ($q$=0) value, 
\bea U_{BF}(\bk)\equiv U_{BF}(k,k,k;ik_n=\bk^2/2m),\eea 
we find
\bea
\lefteqn{\mathrm{Tr}[A^{-1}_0\Lambda^{(2)}] =}&&\nonumber\\&&  
-\sum_{q} c^*_q
c_q\sum_{\bk}U_{BF}(\bk)f(\xi'_{F,\bk})\label{TrA0L2b}
\eea
and
\bea
\lefteqn{\mathrm{Tr}[(A^{-1}_0\Lambda^{(1)})^2] =}&&\nonumber\\&& \sum_{q}\left(c^*_qc_q + c_qc_{-q}  + c^*_qc^*_{-q}+c_{-q}c^*_{-q}\right)\nonumber\\&&\times \sum_{\bk}n_cU^2_{BF}(\bk)\frac{f(\xi'_{F,\bk}) - f(\xi'_{F,\bk+\bq})}{iq_m + \xi'_{F,\bk} - \xi'_{F,\bk+\bq}},
\label{TrA0L1L1b}\eea
where $\xi'_F$ is defined in Eq.~(\ref{A_02}):
\bea \xi'_{F,\bk} = \xi_{F,\bk} + n_cU_{BF}(k,k,k).\label{xiF}\eea
Using the Born approximation value $U^{(0)}_{BB}\equiv U_{BB}(0,0,0)$ for the boson-boson interaction [to facilitate comparison with Eq.~(\ref{S2sp})], the Gaussian fluctuation terms $S^{(2)}_B$ of the Bose action given in Eq.~(\ref{SB}) reduce to
\bea S^{(2)}_B &=& \frac{1}{2}\sum_q\left(-iq_m
+\xi_{B,\bq}\right)c^*_{q}c_{q}\nonumber\\
&&+
\frac{n_c}{2}\sum_{q}U^{(0)}_{BB}\left(c_{q}c_{-q}
+ 4c^*_qc_q + c^*_{-q}c^*_{q}\right).\label{S2B}
\eea  
The requirement that the saddle-point action given in Eq.~(\ref{S0}) be stationary ($\partial S^{(0)}/\partial n_c = 0$) gives 
\bea \mu_B = U^{(0)}_{BB}n_c + \sum_{\bk}U_{BF}(\bk)f(\xi'_{F,\bk}).\label{muB}\eea

Using Eqs.~(\ref{TrA0L2b})-(\ref{muB}), it is straightforward to show that the Gaussian action in Eq.~(\ref{S2}) for the Bose fluctuations of a Bose-Fermi mixture is the same as the Gaussian action of a spin-polarized superfluid given by Eq.~(\ref{S2sp}).  For the Bose-Fermi mixture, the effective boson-boson interaction is then given by the expression in Eq.~(\ref{UBBtilde}) with
\bea U_{BF}(\bk) = \frac{g(\bk)}{n_c}.\label{UBFGauss}\eea 
We note that this expression for $U_{BF}$ has only been derived within a Gaussian approximation.  In Sec.~\ref{quartic}, we show how the inclusion of quartic fluctuations leads to an improved result.  Using Eqs.~(\ref{term}) and (\ref{nc}) as well as $\mu_{\downarrow} = E_b$, Eq.~(\ref{UBFGauss}) reduces to 
\bea U^{(0)}_{BF} \equiv \frac{2\pi (8a_F/3)}{m_r}.\label{UBF0}\eea This corresponds to the Born-approximation value of the dimer-fermion interaction~\cite{Strinati06,deMelo06}, with reduced mass $m_r = 2m/3$.   
Substituting Eq.~(\ref{UBF0}) into Eq.~(\ref{conditionPS}), we find that the stability condition for the spin-polarized Fermi superfluid becomes
\bea \delta n_F^{1/3} \le \frac{(6\pi^2)^{2/3}}{3m}\frac{U^{(0)}_{BB}}{(U^{(0)}_{BF})^2}.\label{conditionPS2}\eea
This requirement is identical to the stability condition for atomic Bose-Fermi mixtures (see, for instance, Eq.~(9) in Ref.~\cite{Pethick00}).

By considering the Gaussian fluctuations in the action for a spin-polarized Fermi superfluid, we have shown how the Bogoliubov excitations of the dimer condensate are affected by the remaining unpaired $\uparrow$ fermions.  At the Gaussian level, however, the renormalized dimer-dimer interaction only involves the Born-approximation value for the dimer-fermion interaction $U_{BF}$ as given in Eq.~(\ref{UBF0}).  To derive a better expression for the dimer-fermion interaction, 
we need to include quartic fluctuation terms.  This is done in Sec.~\ref{quartic}.

\section{Gaussian fluctuations in a spin-polarized  normal Fermi gas}
\label{Gaussiannormal}
In this section, we consider the Gaussian fluctuations of a spin-polarized gas in the normal phase.  The advantage of doing this is that we do not need to work with the small $q$ expansion for the two-fermion vertex function, as we did in the superfluid phase discussed in Sec.~\ref{Bogspectrum}.  We use the results obtained in this Section in our derivation of the effective dimer-fermion interaction given in Sec.~\ref{quartic}.  

Even though we carry out our analysis in the normal state, we restrict ourselves to temperatures well below the dimer binding energy.  Thus, we are still dealing with a Bose-Fermi mixture.  The dimers, however, are not Bose-condensed.  We emphasize, though, that in the extreme BEC limit, $\Delta_0$ is much smaller than the binding energy of the diatomic molecules even at $T=0$~\cite{Engelbrecht}.  Consequently, in this region, the effective interaction is the same whether the molecules (Cooper pairs) are Bose-condensed ($\Delta_0\neq 0$) or not.  

In the normal state, the Gaussian action given in Eq.~(\ref{S2kb}) reduces to
\bea 
S^{(2)} &=& \sum_q\Lambda^*_q\Lambda_q\Gamma^{-1}_{FF}(q),\label{S2kcn}\eea
where the inverse of the two-fermion vertex function $\Gamma_{FF}(q)$ is given by~\cite{deMelo06}
\bea \Gamma^{-1}_{FF}(q) &\equiv&\frac{1}{U} +
\frac{1}{\beta}\sum_{k}G_{0,\downarrow}(k)G_{0,\uparrow}(k+q)\nonumber\\
&=&\frac{1}{U} +
\sum_{\bk}\frac{1-f(\xi_{\bk,\uparrow}) - f(\xi_{\bk-\bq,\downarrow})}{iq_m - \xi_{\bk,\uparrow} - \xi_{\bk-\bq,\downarrow}}.\label{gamma}\eea
In the discussion to follow, we only consider the low-density limit of the unpaired $\uparrow$ fermions and thus we omit the $f(\xi_{\bk,\uparrow})$ term in Eq.~(\ref{gamma}).  Recall that the Fermi distribution factor  $f(\xi_{\bk-\bq,\downarrow})$ for the $\downarrow$ fermions vanishes at low temperatures since $\mu_{\downarrow} \simeq E_b$ is large and negative.  

As before, we use the regularized two-body potential given by Eq.~(\ref{LS}) to replace $1/U$ in terms of $a_F$.  We thus obtain
\bea \lefteqn{\Gamma^{-1}_{FF}(q) =\sum_{\bk}\left[\frac{1}{iq_m - \xi_{\bk,\uparrow} - 
\xi_{\bk-\bq,\downarrow}} + \frac{m}{\bk^2}\right]- \frac{m}{4\pi a_F}}\;\;\;\;\;\;\;\;&&\nonumber\\
&&\!\!\!\!\!\!\!\!\!\!\!\!\!\!\!\!\!\!\!\!=\frac{m}{4\pi a_F}\Bigg\{\frac{4\pi a_F}{m}\!\sum_{\bk}\left[\frac{1}{iq_m\! -\! \xi_{\bk,\uparrow} \!-\! 
\xi_{\bk-\bq,\downarrow}}\! +\! \frac{m}{\bk^2}\right]\!\!-\! 1\!\Bigg\}. \nonumber\\ \label{gamma2b}\eea
The $\bk$-sum in Eq.~(\ref{gamma2b}) can be carried out analytically (assuming $\mu_{\downarrow}<0$) and we find
\bea \lefteqn{\Gamma^{-1}_{FF}(q) =}&&\nonumber\\&&\!\!\!\!\frac{m}{4\pi a_F}\left[\sqrt{\frac{-iq_m + \bq^2/2M -\mu_{\downarrow} -\mu_{\uparrow}}{|E_b|}}-1\right]. \label{gamma3} \eea
After some rearranging, the vertex function in Eq.~(\ref{gamma3}) can be written as
\bea \Gamma_{FF}(\bq,iq_m)= \frac{4\pi}{m^2a_F}\frac{1 + \sqrt{\frac{-iq_m + \xi_{B,\bq} + |E_b|}{|E_b|}}}{-iq_m + \xi_{B,\bq}},\label{gamma2}\eea
where $\xi_{B,\bq} \equiv \bq^2/2M - \mu_{B}$ with the boson chemical potential $\mu_B \equiv \mu_{\downarrow} +  \mu_{\uparrow}+ |E_b|$~\cite{Strinati06,deMelo06}.  For zero polarization, Eq.~(\ref{gamma2}) reduces to the standard particle-particle vertex function $\Gamma$ given in Refs.~\cite{Haussmann, Strinati00}.
In the BEC region, the two-fermion vertex function $\Gamma_{FF}(q)$ in Eq.~(\ref{gamma2}) has a pole arising from bosonic dimer molecules.  

Substituting Eq.~(\ref{gamma2}) into Eq.~(\ref{S2kcn}), the Gaussian action becomes
\bea
S^{(2)} &=& \sum_q\Lambda^*_q\Lambda_q\lambda^{-2}_q\left(-iq_m + \xi_{B,\bq} \right),\label{S2kdn}\eea 
where the strength of the bound state pole in Eq.~(\ref{gamma2}) is
\bea \lambda^2_q &\equiv& \lambda^2(\bq,iq_m)\nonumber\\&&\!\!\!\!\!\!\!\!\!\!\!\!\!\!\!\!\!= \frac{4\pi}{m^2a_F}\left[1+\sqrt{\frac{-iq_m + \xi_{B,\bq} + |E_b|}{|E_b|}}\right].\label{lambda} \eea  
Defining the Bose fields \bea c_q \equiv \lambda^{-1}_q\Lambda_q,\;c^*_q \equiv \lambda^{-1}_q\Lambda^*_q,\label{c}\eea the Gaussian action in Eq.~(\ref{S2kdn}) assumes the more familiar form
\bea
S^{(2)} &=& \sum_{\bq,q_m} \left(-iq_m + \xi_{B,\bq}\right)c^*_q c_q
\nonumber\\
&\equiv&-\sum_q c^*_q D_0^{-1}(q)c_q,\label{S2ken}\eea 
where we have defined the free-space molecular Bose propagator 
\bea D_0(q) \equiv \frac{1}{iq_m - \xi_{B,\bq}}. \label{Dn}\eea

Note that at energies well-below the molecular binding energy (i.e., $iq_m,\xi_{B,\bq}\ll |E_b|$), the Bose fluctuation fields defined in Eq.~(\ref{c}) reduce to $c_q = \sqrt{8\pi/m^2a_F}\Lambda_q$, which is the same field we used in our discussion of the superfluid state, given by Eq.~(\ref{cq}).  Our definition of $c_q$ here represents a very natural separation of the two energy scales in the problem.  The propagator $D_0(q)$ in Eq.~(\ref{Dn}) describes the low-energy dynamics of the dimer molecules through their effective single-particle energy, $\xi_{B,\bq}$.  The momentum- and energy-dependence that enters through $\lambda_q$, in contrast, describes the high-energy internal degrees of freedom of the molecules characterized by the binding energy $E_b$.

\section{Quartic fluctuations}
\label{quartic}
We now consider the quartic fluctuations to the effective Bose actions for a spin-polarized superfluid as well as a Bose-Fermi mixture.  We use this comparison to identify an expression for the effective dimer-unpaired fermion interaction $U_{BF}$ in spin-polarized superfluids that is valid beyond the Born approximation.  To make the physics as clear as possible, we work in the normal state.  

We start by discussing the quartic fluctuations of the effective Bose-Fermi theory.  The quartic fluctuation terms in Eq.~(\ref{SeffB}) are 
\bea S^{(4)} &=& S^{(4)}_B + \frac{1}{2}\mathrm{Tr}[(A^{-1}_0\Lambda^{(2)})^2]  \nonumber\\&&\!\!\!\!\!\!\!\!\!\!\!- \mathrm{Tr}[(A^{-1}_0\Lambda^{(1)})^2A^{-1}_0\Lambda^{(2)}]+ \frac{1}{4}[(A^{-1}_0\Lambda^{(1)})^4]. \label{S4}\eea
In the normal state $\Lambda^{(1)}$ vanishes and Eq.~(\ref{S4}) reduces to
\bea S^{(4)} =  \frac{1}{2\beta}\sum_{q,q',K}V^{(4)}_{BF}(q,q',K)c^*_{\frac{K}{2}+q'}c^*_{\frac{K}{2}-q'}c_{\frac{K}{2}-q}c_{\frac{K}{2}+q}.\nonumber\\ \label{S4BF}\eea
The effective boson-boson interaction $V^{(4)}_{BF}$ in this expression is \bea V^{(4)}_{BF}(q,q',K)=U_{BB}(q,q',K) + \tilde{V}^{(4)}_{BF}(q,q',K),
 \label{V4BF}\eea 
where, using the shorthand $Q^{\pm}\equiv (q\pm q')/2$, 
\bea \lefteqn{\tilde{V}^{(4)}_{BF}(q,q',K) \equiv}&&\nonumber\\&& 
\frac{1}{\beta}\sum_{k} U_{BF}\Bigg(k+Q^{+}\! +\!\frac{K}{4},
k-Q^{+}\!+\!\frac{K}{4},k +\!\frac{3K}{4}\!+\!Q^{-}\!\! \Bigg)\nonumber\\&&
\times U_{BF}\Bigg(k-Q^{+}\!+\!\frac{K}{4},
k+Q^{+}\!+\!\frac{K}{4},k+\!\frac{3K}{4}\!-\!Q^{-}\Bigg)\nonumber\\&&
\times A^{-1}_{0}\left(k + Q^{+}\!+\!\frac{K}{4}\right)
A^{-1}_{0}\left(k - Q^{+}\!+\!\frac{K}{4}\right).\label{V1BF}\eea  
In this expression, the $A^{-1}$ Green's functions are given by their $n_c=0$ value in Eq.~(\ref{A_02}): 
\bea A^{-1}_0(k) &=& \frac{1}{ik_n - \xi_{F,\bk}}.\label{A_03}\eea


Taken together, Eqs.~(\ref{V4BF}) and (\ref{V1BF}) describe the leading-order terms in the effective boson-boson interaction in a Bose-Fermi mixture.  The first term in Eq.~(\ref{V4BF}) is the bare boson-boson interaction $U_{BB}$ while the second term, $\tilde{V}^{(4)}_{BF}$, given by Eq.~(\ref{V1BF}), describes an effective boson-boson interaction mediated by the fermions.  We will show that the quartic fluctuation action in Eq.~(\ref{S4BF}) has the same form as the quartic fluctuation action for a spin-polarized superfluid.

The quartic fluctuation terms in the action for a spin-polarized gas given by Eq.~(\ref{SeffQ}) are 
\bea \lefteqn{S^{(4)} = \frac{1}{4\beta^2}\mathrm{Tr}[(\bG_0\bSig)^4]=}&&\nonumber\\&& \!\!\!\!\!\!\!\!\!\!\!\frac{1}{2\beta}\sum_{q,q',K}V^{(4)}(q,q',K)c^*_{\frac{K}{2}+q'}c^*_{\frac{K}{2}-q'}c_{\frac{K}{2}-q}c_{\frac{K}{2}+q}. \label{S4c}\eea
Here, we have used the Bose field $c_q$ defined in Eq.~(\ref{c}).  As before, $K\equiv (\bK,iK_m)$ is a four-vector denoting the momentum $\bK$ and Bose Matsubara frequency $iK_m$, corresponding to the centre-of-mass degree of freedom of two bosons.  The interaction $V^{(4)}$ in Eq.~(\ref{S4c}) is found to be
\bea \lefteqn{V^{(4)}(q,q',K) =}&&\nonumber\\&& \lambda_{K/2+q'}\lambda_{K/2-q'}\lambda_{K/2-q}\lambda_{K/2+q}
\frac{1}{\beta}\sum_k\nonumber\\&& G_{0,\downarrow}\left(k + Q^{-} - \frac{K}{4}\right)G_{0,\downarrow}\left(k -Q^{-} - \frac{K}{4}\right)\nonumber\\
&& \times G_{0,\uparrow}\left(k + Q^{+} + \frac{K}{4}\right)G_{0,\uparrow}\left(k -Q^{+} + \frac{K}{4}\right). \nonumber\\ \label{V1F}\eea  
In the case of an unpolarized Fermi gas, this expression reduces to the interaction discussed by Ohashi~\cite{Ohashi05} when we approximate $\lambda_q$ by $\lambda_0$.   

Since a spin-polarized Fermi superfluid in the BEC region is equivalent to a Bose-Fermi mixture, when $\mu_{\uparrow}\neq \mu_{\downarrow}$ (i.e., $n_{\uparrow}\neq n_{\downarrow}$), the effective interaction given in Eq.~(\ref{V1F}) includes an effective dimer-dimer interaction mediated by the gas of unpaired $\uparrow$ fermions.  In addition, Eq.~(\ref{V1F}) includes a direct dimer-dimer interaction which, unlike the mediated interaction, does not vanish when $n_{\uparrow} = n_{\downarrow}$.   However, unlike the expression given in Eq.~(\ref{V4BF}) for the effective boson-boson interaction 
in a Bose-Fermi mixture, in Eq.~(\ref{V1F}) these two different contributions are not explicitly separated.  We will show that the mediated interaction in Eq.~(\ref{V1F}) is given by the poles of the $G_{0,\uparrow}$ for the unpaired $\uparrow$ fermions.  The direct contribution is given by the poles of the Green's functions $G_{0,\downarrow}$ for the paired fermions.  We shall use the notation $\sum_{G_{0,\sigma}}$ to denote the contribution to the Fermi Matsubara frequency sum from the residue of the poles of the $G_{0,\uparrow}$ and $G_{0,\downarrow}$ Green's functions in Eq.~(\ref{V1F}).  Using this, the direct and mediated interactions can be separated as follows:                        
\bea V^{(4)}(q,q',K)=V^{(4)}_{\mathrm{direct}}(q,q',K) + V^{(4)}_{\mathrm{mediated}}(q,q',K),\nonumber\\ \label{VseparateF}\eea
where  \bea V^{(4)}_{\mathrm{direct}}(q,q',K) &\equiv& \sum_{\bk}\sum_{G_{0,\downarrow}}\left\{\;\cdots\;\right\},\label{Vdir}\eea
\bea V^{(4)}_{\mathrm{mediated}}(q,q',K) &\equiv& \sum_{\bk}\sum_{G_{0,\uparrow}}\left\{\;\cdots\;\right\},\label{Vmed}\eea
and $\{\;\cdots\;\}$ denotes the integrand in Eq.~(\ref{V1F}).

We now show that $V^{(4)}_{\mathrm{direct}}$ can be identified as a ``direct" dimer-dimer interaction, whereas $V^{(4)}_{\mathrm{mediated}}$ is an interaction between dimer molecules that is mediated by unpaired fermions.  Recalling that the chemical potential $\mu_{\uparrow}$ for the majority species of fermions is just the chemical potential for a gas of the excess unpaired fermions, we see that $A^{-1}_{0}(k) = G_{0,\uparrow}(k)$, where $A^{-1}_0(k)$ is given by Eq.~(\ref{A_03}) and  $G_{0,\uparrow}(k)$ is given by Eq.~(\ref{Gup}).  Using this identification to compare Eqs.~(\ref{Vmed}) and (\ref{V1BF}) leads to the result
\bea U_{BF}(k,k',P) &=& \lambda_{P\!-\!k}G_{0,\downarrow}(k\!+\!k'\!-\!P)\lambda_{P\!-\!k'},\label{UBF2}\eea
where $\lambda_q$ is defined in Eq.~(\ref{lambda}) and $k,k',$ and $P$ all involve Fermi frequencies.  This is a key result of the present paper.    

Note that in making the identification in Eq.~(\ref{UBF2}), one must assume that the poles of the frequency-dependent boson-fermion interaction $U_{BF}$ in Eq.~(\ref{V1BF}) cannot contribute to the frequency sum in that expression.  This would of course be the case for a boson-fermion interaction between ``elementary" bosons and fermions, in which case $U_{BF}$ would be frequency-independent.  In the present context, however, where $U_{BF}$ must be frequency-dependent owing to the fact that the bosons are dimer molecules, we must impose this restriction on the frequency sum in the term describing interactions between bosons and fermions in the Bose-Fermi model introduced in Eq.~(\ref{Seff}).     

Following the identification of Eq.~(\ref{Vmed}) with Eq.~(\ref{V1BF}), one sees that the ``direct" dimer-dimer interaction given by Eq.~(\ref{Vdir}) is equivalent to $U_{BB}$ in Eq.~(\ref{V4BF}), 
\bea U_{BB}(q,q',K) = V^{(4)}_{\mathrm{direct}}(q,q',K).\label{UBB3}\eea

We now consider what happens to the direct and mediated interactions identified in Eqs.~(\ref{Vdir}) and (\ref{Vmed}) as the polarization vanishes, that is, in the limit where $\mu_{\downarrow} = \mu_{\uparrow} = E_b/2$.  At $T=0$ in the BEC region, all fermions are paired up to form dimers so there are no unpaired fermions and the mediated interaction $V^{(4)}_{\mathrm{mediated}}$ in Eq.~(\ref{Vmed}) vanishes as expected.  On the other hand, at finite temperatures $V^{(4)}_{\mathrm{mediated}}$ does not vanish.  Even in the absence of polarization, there is a dimer-dimer interaction mediated by unpaired fermions since some of the dimer molecules will be thermally dissociated.  As long as $|\mu| \gg T$, however, the number of dissociated molecules is small in an unpolarized gas, even close to the transition temperature $T_c$~\cite{deMelo}.  Thus, in the BEC region of strongly-bound dimer molecules, the mediated interaction in Eq.~(\ref{Vmed}) is only significant in a spin-polarized gas.  In contrast, the direct interaction given in Eq.~(\ref{Vdir}) remains finite even for zero polarization.  At $T=0$, one can show that (using $\mu_{\downarrow} = -1/ma^2_F$)\bea V^{(4)}_{\mathrm{direct}}(0,0,0) &=& \left(\frac{8\pi}{m^2a_F}\right)^{2}\sum_{\bk}\frac{2}{\left(\bk^2/m + |\mu_{\downarrow}|\right)^3} \nonumber\\&=& \frac{4\pi(2a_F)}{M}.\label{VdirBorn}\eea This reproduces the Born-approximation value for the direct dimer-dimer interaction given in Eq.~(\ref{UBB0}).   This confirms that Eq.~(\ref{Vdir}) is the direct interaction between dimer molecules, i.e., the part of the interaction that is not mediated by the unpaired excess $\uparrow$ fermions.

To summarize, we have compared the quartic fluctuation terms in the action for a spin-polarized Fermi gas with those for the action given in Eq.~(\ref{Seff}) which describes a Bose-Fermi mixture.  We have thus proven that a spin-polarized Fermi gas in the BEC limit reduces to a Bose-Fermi mixture described by Eq.~(\ref{Seff}).  An important caveat is that the poles of $U_{BF}$ [given in Eq.~(\ref{UBF2})] are not allowed to contribute to the frequency sum in the dimer-fermion interaction term in Eq.~(\ref{Seff}).


In closing this Section, we discuss the expression for $g(\bk)$ in Eq.~(\ref{term}).  With the expression for $U_{BF}$ given in Eq.~(\ref{UBF2}), we can rewrite $g(\bk)$ as (in the limit where $\bk^2/2m \ll |E_b|$)
\bea g(\bk) &\simeq&\left(\frac{\Delta^2_0 m^2a_F}{8\pi}\right)\frac{8\pi}{m^2 a_F}G_{0,\downarrow}(\bk;ik_n=\bk^2/2m)\nonumber\\
&=&n_cU_{BF}(k,k,k;\Delta_0=0;ik_n=\bk^2/2m).\label{gk}\eea
Note that $g(\bk)$ arises in the effective theory of a spin-polarized Fermi superfluid after performing an expansion in powers of $\Delta_0$, treating $\mu_{\uparrow}\ll |\mu_{\downarrow}|$ [see Eq.~(\ref{UBBtilde})].  Thus, it is not surprising that $g(\bk)$ is given by the free-space value of the dimer-fermion interaction $U_{BF}$, with $n_c = \delta n_F = 0$, (i.e., $\Delta_0=0$ and $\mu_{\uparrow}=0$) and with the corresponding energy $ik_n = \bk^2/2m$ for a free-space fermion.

\section{The dimer-fermion scattering length}
\label{TBSA}

As a simple application of our expression for the dimer-fermion scattering interaction in a spin-polarized Fermi gas given by Eq.~(\ref{UBF2}), we discuss the scattering properties of a dimer-fermion pair in free-space.  This allows us to make contact with the extensive literature discussing this problem~\cite{Brodsky05,Brodsky05PRA,Gurarie06,Bedaque98} and the well-known exact result $a_{BF}=1.18a_F$ due to Skorniakov and Ter-Martirosian~\cite{Skorniakov57} for the dimer-fermion scattering length.  

\begin{figure}
\begin{center}
\epsfig{file=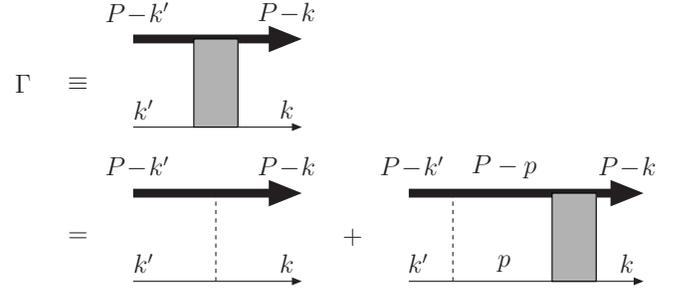, angle=0,width=8.5cm}
\caption{The Bethe-Salpeter equation for the renormalized low-energy dimer-fermion interaction $\Gamma$ within the ladder approximation, Eq.~(\ref{BS}).  The thick lines denote the Bose propagator $D_0$ while the thin lines refer to the propagator $G_{0,\uparrow}$ for unpaired fermions. The dashed line is the dimer-fermion interaction given by Eq.~(\ref{UBF3}).}
\label{BSdiagram}
\end{center}
\end{figure}

For free-space scattering of a single dimer-fermion pair, we can set $\mu_{\uparrow}=0$, and $|\mu_{\downarrow}| = |E_b|$ in Eq.~(\ref{UBF2}) to give
\bea \lefteqn{U_{BF}(k,k',P) = \frac{4\pi a_F/m}{ik_n + ik'_n - iP_n + \epsilon_{F,\bk\!+\!\bk'\!-\!\bP} + |E_b|}\times}&&\nonumber\\&&
\left[\sqrt{|E_b|} + \sqrt{-iP_n +ik_n + \epsilon_{B,\bP\!-\!\bk} + |E_b|}\right]^{1/2}\times\nonumber\\&&\left[\sqrt{|E_b|} + \sqrt{-iP_n +ik'_n +\epsilon_{B,\bP\!-\!\bk'} + |E_b|}\right]^{1/2},\nonumber\\ \label{UBF3}\eea   
where, $\epsilon_{F,\bk}\equiv \bk^2/2m$ and $\epsilon_{B,\bk}\equiv \bk^2/2M = \bk^2/4m$.  
In order to calculate the $s$-wave dimer-fermion scattering amplitude associated with the interaction given in Eq.~(\ref{UBF3}), we need to calculate a renormalized low-energy interaction $\Gamma$ which is obtained from the ``bare" interaction $U_{BF}$ given by Eq.~(\ref{UBF3}) by integrating out short-wavelength (i.e., smaller than $a_F$) degrees of freedom.  Following the usual prescription, $\Gamma$ is given by the Bethe-Salpeter equation (within the ladder approximation~\cite{FW}) for dimer-fermion scattering (see Fig.~\ref{BSdiagram}):
\bea \lefteqn{\Gamma(P-k',k';P-k,k) =U_{BF}(k,k',P)}&&\nonumber\\&-&\frac{1}{\beta}\sum_{\bp,p_n}U_{BF}(p,k',P) D_0(P\!-\!p)G_{0,\uparrow}(p)\nonumber\\ &&\times\Gamma(P\!-\!p,p;P\!-\!k,k).\label{BS}\eea   
$D_0$ is the free-space molecular propagator for the dimers, 
\bea D_0(q) = \frac{1}{iq_m - \epsilon_{B,\bq}},\eea and $G_{0,\uparrow} = (ik_n - \epsilon_{F,\bk})^{-1}$ is the propagator for unpaired $\uparrow$ fermions in free-space.  The sum in Eq.~(\ref{BS}) is over the Fermi Matsubara frequency $p_n$.  
Consistent with our discussion in Sec.~\ref{quartic}, the poles of the Green's function $G_{0,\downarrow}$ involved with $U_{BF}$ do not contribute to the frequency sum over $p_n$ [see Eq.~(\ref{BS2})].

The $s$-wave boson-fermion scattering length $a_{BF}$ is given in the usual way by (see also Refs.~\cite{Brodsky05,Brodsky05PRA,Gurarie06})
\bea a_{BF} \equiv \frac{m_r}{2\pi}\Gamma(0,0;0,0),\label{aBFdef}\eea
where $m_r$ is defined below Eq.~(\ref{UBF0}).
To calculate $a_{BF}$, we need only consider $\Gamma(-k',k';0,0)$, i.e., we set the outgoing fermion energy and momentum $k$ equal to zero as well as the centre-of-mass energy and momentum $P$ (see Fig.~\ref{BSdiagram}).  In addition, for scattering in free space, we can use the on-shell energy value $ik'_n = \epsilon_{F,\bk'}$.  Using these,  Eq.~(\ref{BS}) reduces to 
\bea \lefteqn{\Gamma(-\bk',\bk') =\frac{\lambda_0\lambda_{-\bk'}}{2\epsilon_{F,\bk'} + |E_b|}}&&\nonumber\\&+&\frac{1}{\beta}\sum_{\bp,p_n}\frac{\lambda_{-p}\lambda_{-\bk'}}{\left(ip_n+\epsilon_{F,\bk'} + \epsilon_{F,\bp\!+\!\bk'}+|E_b|\right)}\nonumber\\&&
\times\frac{\Gamma(-p,p,0,0)}{\left(ip_n + \epsilon_{B,\bp}\right)\left(ip_n-\epsilon_{F,\bp}\right)},\label{BS2}\eea 
where we have defined $\Gamma(-\bk',\bk')\equiv \Gamma(-k',k';0,0;ik'_n = \epsilon_{F,\bk'})$ and
\bea \lambda_{\bk} &\equiv& \lambda(\bk,ik_n=\epsilon_{F,\bk}) \nonumber\\ &=& \left[\frac{4\pi }{m^2a_F}\left(1 + \sqrt{\left(3\bk^2/4m + |E_b|\right)/|E_b|}\right)\right]^{1/2}.\nonumber\\ \eea

Since we are dealing with the limit of vanishing density of unpaired fermions, we work at $T=0$.  Analytically continuing the imaginary Fermi frequency $ip_n$ to the real frequency $\omega_p$, and carrying out the integration over $\omega_p$,  Eq.~(\ref{BS2}) reduces to
\bea   \lefteqn{\Gamma(-\bk',\bk') =\frac{\lambda_{0}\lambda_{-\bk'}}{2\epsilon_{F,\bk'} + |E_b|}}&&\nonumber\\&-&\sum_{\bp}\frac{\lambda_{-\bp}\lambda_{-\bk'}}{\left(\epsilon_{F,\bp}+\epsilon_{F,\bk'} + \epsilon_{F,\bp\!+\!\bk'}+|E_b|\right)}\frac{1}{\left(\epsilon_{F,\bp} + \epsilon_{B,\bp}\right)}\nonumber\\&&\times\Gamma(-\bp,\bp).\label{BStilde5}\eea  
Equation~(\ref{BStilde5}) is an integral equation for the momentum-dependent dimer-fermion scattering vertex $\Gamma$.  Following the definition used in Ref.~\cite{Brodsky05PRA}, we define the scattering amplitude as
\bea a_{BF}(\bk') \equiv \frac{m_r}{2\pi}\Gamma(-\bk',\bk')\frac{\lambda_{-\bk'}}{\lambda_0}.\label{aBFkn}\eea The $s$-wave dimer-fermion scattering length $a_{BF}$ defined in Eq.~(\ref{aBFdef}) corresponds to $a_{BF}(\bk'=\b0)$.  One can show that 
Eq.~(\ref{BStilde5}) leads to the following integral equation for $a_{BF}(\bk')$:
\bea \lefteqn{\frac{3a_{BF}(\bk')/4}{\sqrt{m|E_b|} + \sqrt{3\bk'^2/4 + m|E_b|}} = \frac{1}{\bk'^2 + m|E_b|}}&&\nonumber\\&-&4\pi\sum_{\bp}\frac{a_{BF}(\bp)}{\bp^2\left(\bp^2 + \bp\cdot\bk' + \bk'^2+m|E_b|\right)}. \label{aBFvacuo}\eea 
This is precisely the Skorniakov-Ter-Martirosian integral equation for the scattering amplitude~\cite{Skorniakov57,Brodsky05,Brodsky05PRA,Gurarie06}.  Solving this integral equation for $a_{BF}(\bk')$, one finds $a_{BF}(\bk'=\b0)\equiv a_{BF}=1.18a_F$.

\section{Conclusions}  
In the BEC limit of a spin-polarized Fermi superfluid, fermions in different hyperfine states pair up to form dimer molecules.  When the binding energy of these molecules is larger than any other energy scale in the problem (i.e., $|E_b|\gg\Delta_0, T,\mu_{\uparrow}$) the spin-polarized Fermi superfluid reduces to a Bose-Fermi mixture, where the fermions are the unpaired excess fermions with chemical potential $\mu_{\uparrow}$.  Using a functional integral approach and comparing the fluctuation terms in the effective Bose actions of a Bose-Fermi mixture and spin-polarized Fermi superfluid, we have shown precisely how this equivalence emerges in the BEC region. 

At the Gaussian level, the fluctuations of the spin-polarized superfluid are described by the usual BEC Bogoliubov propagator in Eq.~(\ref{S2sp}).  The new dimer-dimer interaction in Eq.~(\ref{UBBtilde}) includes a contribution that is mediated by the unpaired fermions and is described by a Lindhard response function. 

Going beyond the Gaussian level by including quartic fluctuations about the mean-field, we derived an explicit expression in Eq.~(\ref{UBF2}) for the interaction between the dimers and unpaired fermions in the spin-polarized Fermi gas.  Taking the free-space value of this interaction, we showed how our results reproduce the well-known value for the $s$-wave scattering length $a_{BF}=1.18a_F$, first derived in 1957~\cite{Skorniakov57}. More recently this same result has been derived for three interacting fermions using a diagrammatic approach by Brodsky and co-workers~\cite{Brodsky05,Brodsky05PRA} as well as by Levinsen and Gurarie~\cite{Gurarie06}.

\begin{acknowledgments}
E.T. thanks Dr.~Erhai Zhao for helpful discussions.  E.T. and A.G. were supported by NSERC of Canada.   Y.O. was
financially supported by Grant-in-Aid
for Scientific research from the Ministry of Education, Culture, Sports,
Science and Technology of Japan and CREST(JST).
\end{acknowledgments} 

\appendix


\begin{thebibliography}{99}
\bibitem{note} In the BEC limit and for small polarizations, $\Delta_0\sim |E_b|(n_Fa^3_F)^{1/2}$~\cite{Engelbrecht}, where $n_F$ is the total number of fermions.   Thus, $(\Delta_0/|E_b|)^2 \sim {\cal{O}}[n_F a^3_F]$.
\end{thebibliography}
\end{document}